\documentclass[aps, prl, twocolumn,a4paper,superscriptaddress,floatfix,showpacs]{revtex4-2}

\usepackage[paperwidth=215mm,paperheight=300mm,centering,hmargin=1.6cm,vmargin=2cm]{geometry}
\usepackage{graphicx,amsmath,SIunits}
\usepackage{latexsym,stmaryrd}
\usepackage[dvipsnames]{xcolor}
\usepackage[normalem]{ulem}
 \usepackage{amsmath}
 \usepackage{array}
\usepackage{xspace}
\makeatletter

\newcommand\alexout{\bgroup\markoverwith{\textcolor{blue}{\rule[0.5ex]{1pt}{1pt}}}\ULon}

\begin{document}

\title{Boosted Optomechanics with a Fluid of Nonlinear Polaritons}

\author{Florent Malabat}
\affiliation{Matériaux et Phénomènes Quantiques, Université Paris Cité, CNRS, Paris, France}

\author{Martin Colombano}
\affiliation{Matériaux et Phénomènes Quantiques, Université Paris Cité, CNRS, Paris, France}

\author{Titouan Lévêque}
\affiliation{Matériaux et Phénomènes Quantiques, Université Paris Cité, CNRS, Paris, France}

\author{Martina Morassi}
\affiliation{Centre de Nanosciences et Nanotechnologies, Université Paris-Saclay, CNRS, Palaiseau, France}

\author{Aristide Lema\^itre}
\affiliation{Centre de Nanosciences et Nanotechnologies, Université Paris-Saclay, CNRS, Palaiseau, France}

\author{Alexandre Le Boit\'e}
\affiliation{Matériaux et Phénomènes Quantiques, Université Paris Cité, CNRS, Paris, France}

\author{Ivan Favero}
\email{ivan.favero@u-paris.fr}
\affiliation{Matériaux et Phénomènes Quantiques, Université Paris Cité, CNRS, Paris, France}

\date{\today}

\small

\begin{abstract}

Merging optomechanics and polaritonics opens stimulating perspectives like the giant enhancement of optomechanical interaction and the enrichment of optomechanics with effective nonlinear photons. The experimental implementation of these concepts has however remained elusive. Here we report on the resonant optical control of polaritonic optomechanical resonators constituted of semiconductor disks embedding quantum wells. Whispering gallery photons and quantum well excitons strongly couple, leading to the emergence of polaritons that couple to the mechanical vibrations of the disk. We perform resonant optomechanical frequency response experiments on these resonators, modeled introducing a minimal set of constitutive equations, from which we extract the polariton-modified optomechanical coupling $g_0$ and the polariton nonlinearities. We observe a boost of $g_0$ by more than a decade compared to bare photons, reaching to a record $g_0$ for whispering gallery resonators of $22$ MHz, and analyze experimentally and theoretically  its evolution as function of the polariton's composition. We also measure a clear hierarchy of three polaritonic nonlinearities, again analyzed as function of polariton composition, establishing a bridge between past unconciliated reports in polaritonics. Grounded on experimental and theoretical foundations, resonant polaritonic optomechanics is set ready for an optomechanical exploration of quantum fluids of polaritons.

\end{abstract}

%\pacs{(42.25.Dd, 42.25.Fx, 46.65.+g, 42.70.Qs)}

\maketitle

\section{Introduction}

Cavity optomechanics studies the interaction between light and vibrating mechanical bodies \cite{Favero2009,CavityOptomechanics2014}, with applications in the transduction of quantum signals \cite{Stockill2024,Mirhosseini2023}, in sensing \cite{Sansa2020,Sbarra2022}, or in many-body phenomena \cite{KitaevChain2024,GilSantos2017}. The optomechanical interaction is parametrized by the vacuum coupling $g_0$ between a single photon and a single phonon, whose enhancement has constituted a long-standing goal \cite{Rabl2010}. Rapid progress on $g_0$ was obtained in the early days of the field but the pace has recently slowed down, interrogating our capability to further tailor photon-phonon interactions. Beyond the canonical case of radiation pressure \cite{Law}, several physical processes can produce optomechanical coupling. Electrostriction is a conservative effect that occurs without absorption \cite{Rakich2010,Chan2012PhotoElastic,Favero2014} and increases when approaching electronic resonances of a material \cite{Feldman1968}. Light can be absorbed and produce thermal strain \cite{metzger2008optical,Restrepo2011}, a dissipative effect dubbed bolometric or photothermal. Light resonant with electronic transitions can be employed to excite free or confined carriers, which generate in turn strain via piezoelectric \cite{Yamaguchi2011} or deformation potential \cite{Polzik2018} coupling. This optomechanical effect is potentially large but again non-conservative, as carrier dissipation is generally fast. 

The strong coupling between light and carriers offers an interesting avenue in this context \cite{PolaritonPath2014}. The resulting hybrid light-matter quasiparticles -- the polaritons -- can indeed interact conservatively with mechanical vibrations, while keeping advantage of intense carrier-phonon interaction. By replacing bare photons with polaritons, one could hence obtain an optomechanics with increased coupling $g_0$. At the same time, this approach introduces the ingredient of polaritonic nonlinearity stemming from the polaritons' matter component, leading to an optomechanics with effective nonlinear photons. Polariton-polariton interactions can generate many-body effects in optics, and they greatly contributed to the study of quantum gases of photons, which revisited the concepts of superfluidity, thermalization and Bose-Einstein condensation \cite{PolaritonCondensate2006,Ciuti2013,Weitz2010,Wouters2022}. By merging optomechanics and polaritonics, we hence have the dual opportunity to improve optomechanical coupling, but also to optomechanically explore the nonlinearities, hydrodynamics and thermodynamics of polaritonic quantum fluids of light. The present work precisely seizes this dual opportunity.

Early studies did motivate a convergence between optomechanics and polaritonics. Brillouin scattering experiments carried on quantum wells within a bulk material observed an increase of the scattered signal in the bulk polariton regime \cite{Jusserand2015CavityOptomechanicsPolaritonMQW}. Brillouin scattering and photoluminescence on a planar optical cavity embedding quantum wells showed an intense coupling of cavity polaritons to bulk phonons \cite{Sesin2023GiantOptomechanicalCouplingPolariton}. In the meantime, the quantum theory of interactions in a polaritonic mechanical resonator was investigated \cite{Restrepo2014SinglePolariton,Kyriienko2014}, establishing that the resulting Hamiltonian indeed resembles that of optomechanics but with a boosted $g_0$ and effective photon nonlinearities \cite{CarlonZambon2022}. More recently, non resonant photoluminescence experiments on patch cavities reported locking between polaritons and phonons \cite{Chafatinos2023} and the existence of phonoritons \cite{Kuznetsov2023}, but no direct measurement of $g_0$ nor nonlinearities. In optomechanics, the optical cavity is generally driven at or near resonance in order to produce a coherent response. This condition is a requisite to obtain key effects such as sideband cooling, amplification, measurement at the quantum limit \cite{CavityOptomechanics2014}, and it is also necessary to implement reference methods to measure $g_0$. In polaritonics, resonant experiments were also employed to reveal important phenomena such as superfluidity \cite{Amo2009}, dark solitons \cite{Maitre2020}, or the nucleation of vortices \cite{Nardin2011, Ma2026}. In order to explore the intersection of optomechanics and polaritonics in the richest regimes, resonant coherent control experiments are hence required. They have not been achieved yet on a polaritonic optomechanical resonator. 

Here we report resonant coherent control experiments on polaritonic optomechanical resonators constituted of suspended Gallium Arsenide (GaAs) disks embedding Indium Gallium Arsenide (InGaAs) quantum wells. In a sub-micron volume, these circular resonators confine photons, excitons, and phonons. Photons and quantum-well excitons strongly couple to form whispering gallery exciton-polaritons \cite{DeOliveira2024}, which interact with the breathing mechanical motion of the disk \cite{CarlonZambon2022}. These ingredients give birth to a quantum fluid of polaritons within an optomechanical resonator, or formally an optomechanics with nonlinear polaritons. We present constitutive equations for such a system under resonant laser driving, enabling modeling of its linear frequency response, which we then measure by resonant modulation/demodulation approach \cite{metzger2008optical} analog to optomechanically-induced transparency (OMIT) \cite{Weis2010,Safavi-Naeini2011}. This method allows a calibrated measurement of $g_0$, which exhibits a clear boost that grows with the excitonic character of polaritons. The boost attains a factor $15$ compared to bare photons, corresponding to $g_0=22$ MHz, a record value for a whispering gallery optomechanical resonator. The boost is reproduced by our analytical and numerical calculations, confirming the role played by exciton-phonon couplings and the deformation potential. At the same time, the experimental method allows for a calibrated measurement of polaritonic nonlinearities. Our experiments reveal a clear hierarchy of three distinct polaritonic nonlinearities in the fluid, each associated to a distinct timescale, reconciling the data spread found in the polariton literature. The fluid of polaritons appears to be governed not only by direct Coulomb nonlinearities, but also by thermal and exciton reservoir-mediated nonlinearities. With the $g_0$ and nonlinearities measured and modeled, we reach to a complete control of boosted optomechanics with polaritons. The dynamics of the polaritonic fluid coupled to mechanical motion is perfectly reproduced by our constitutive equations, which lays the foundations for an optomechanical exploration of fluids of light.

\begin{figure}[ht]
\includegraphics[width=250pt]{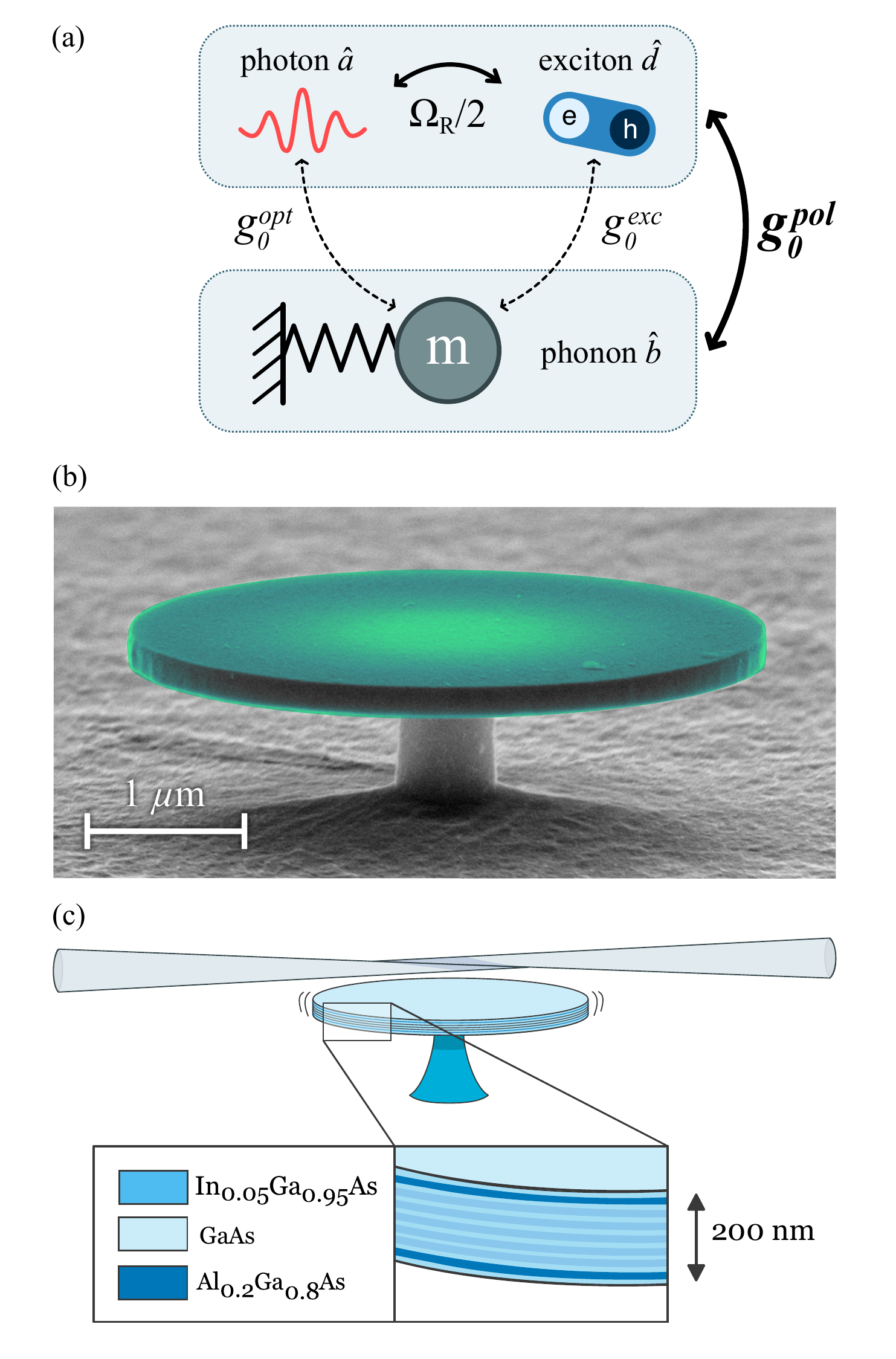}
\caption{\textbf{Principles and resonators}. \textbf{(a)} The mechanical mode $\hat{b}$ couples with the photon mode $\hat{a}$ at a rate $g_0^{opt}$ and with the exciton mode $\hat{d}$ at a rate $g_0^{exc}$. In the strong photon-exciton coupling regime, a polariton $\hat{p}$ is obtained by photon-exciton hybridization, characterized by the Rabi frequency $\Omega_R$. The polariton couples to the mechanical mode at a rate $g_0^{pol}$, which is a linear combination of $g_0^{opt}$ and $g_0^{exc}$. \textbf{(b)} Scanning electron microscope image of a polaritonic mechanical disk of 2 $\mu$m radius and 200 nm thickness. \textbf{(c)} Two homemade conical fibers are joined to form a junction that evanescent couples to the disk resonator. The disk embeds five 12 nm thick ${\rm In}_{0.05}{\rm Ga}_{0.95}{\rm As}$ quantum wells separated by 15 nm GaAs layers. Two ${\rm Al}_{0.2}{\rm Ga}_{0.8}{\rm As}$ layers at the top and the bottom of the structure facilitate exciton trapping, while two GaAs layers at the top and bottom surfaces of the disk protect the AlGaAs layers from oxidation. The pedestal is made of ${\rm Al}_{0.2}{\rm Ga}_{0.8}{\rm As}$, obtained by selective wet etching.}
\label{fig:presentation}
\end{figure}

\section{Principles and model}

The principle of polariton-mechanical interaction is depicted in Fig. 1a. In cavity optomechanics, a photon mode $\hat{a}$ couples to a phonon mode $\hat{b}$ through the parametric interaction $\hat{H}_{int}\slash\hbar = -g_{0}^{opt} \hat{a}^{\dagger} \hat{a} (\hat{b}^\dagger+\hat{b})$. In a semiconductor, an exciton mode $\hat{d}$ couples to a vibration through the deformation potential, adopting a similar interaction form $\hat{H}_{int}\slash\hbar = -g_{0}^{exc} \hat{d}^{\dagger} \hat{d} (\hat{b}^\dagger+\hat{b})$. The strong coupling between quasi-resonant photons and excitons leads to the formation of exciton polaritons. At exciton-photon resonance the eigenstates of these hybrid particles, dubbed lower polaritons (LP) and upper polaritons (UP), are split by the Rabi energy $\hbar\Omega_R$. Focusing on the LP branch, for a mechanical frequency $\Omega_m \ll \Omega_R$, the Hamiltonian of the photon-exciton-phonon system in the LP basis simplifies to \cite{CarlonZambon2022}:
\begin{equation}
\hat{H}\slash\hbar = \omega_l\hat{p}_l^\dagger \hat{p}_l + \Omega_m \hat{b}^\dagger \hat{b} - g_{0}^{lp} \hat{p}_l^{\dagger} \hat{p}_l (\hat{b}^\dagger+\hat{b})+ ({g}/{2}) \hat{p}_l^{\dagger}\hat{p}_l^{\dagger} \hat{p}_l  \hat{p}_l 
\label{Hamiltonian}
\end{equation}
with $\hat{p}_l$ and $\omega_l$ the LP operator and frequency, and $g$ the nonlinear polariton-polariton interaction constant. This is in essence a canonical optomechanical Hamiltonian, with photons replaced by polaritons and with a nonlinear interaction between polaritons. The lower polariton being a linear superposition of photon and exciton $\hat{p}_l =-C \hat{a}+X \hat{d}$, with $C$ and $X$ the Hopefield coefficients \cite{Hopfield1958}, the polariton-phonon coupling $g_{0}^{lp}$ is itself a mix of the optomechanical and exciton-phonon interactions \cite{CarlonZambon2022}: 
\begin{equation}
g_{0}^{lp} = C^2 g_0^\text{opt} + X^2 g_0^\text{exc}
\label{eq:g_0pol}
\end{equation}
where $X^2+C^2=1$ and $X^2=1\slash2\left(1+\delta\slash\sqrt{\delta^2 +\Omega_R^2} \right)$, with $\delta$ the cavity-exciton detuning $E_c - E_x$. The new optomechanical coupling $g_{0}^{lp}$ is hence comprised between \(g_0^\text{opt}\) and \(g_0^\text{exc}\). In the resonator structures studied below, the exciton-phonon coupling \(g_0^\text{exc}\) can be a 100 times larger than \(g_0^\text{opt}\) \cite{CarlonZambon2022}, hence a polaritonic boost of the optomechanical coupling of up to 100 can be expected in principle. In order to approach this regime, the polariton must have a strong excitonic weight X. 

In this work, we aim to measure the polaritonic boost of $g_{0}$ in an unambiguous and calibrated manner. In canonical optomechanics, assessing $g_{0}$ is best achieved by measuring the coherent linear optical response of the system under modulation of the input light and demodulation of the output light \cite{metzger2008optical,Weis2010,Safavi-Naeini2011}. Provided the optical/mechanical frequencies $\omega_{\text{cav}}$/$\Omega_{m}$ and dissipation rates $\kappa$/$\Gamma_{m}$ are measured and the intracavity photon number calibrated, this response leads to a unique value for $g_{0}$. In the present work, photons are replaced by polaritons, hence polariton nonlinearities must be included into the model in order to retrieve the full response of the system. Starting from Eq.~\eqref{Hamiltonian}, we adopt a semiclassical approach that, for the polariton field, is a single-mode equivalent of the dissipative Gross-Pitaevskii equation commonly used to describe exciton-polariton condensates~\cite{Wouters2007}.  Adding drive and losses and moving to a frame rotating at the laser frequency we obtain two coupled equations for the polariton and mechanical fields:

\begin{equation}\label{omPol}
\begin{gathered} 
\dot{p}=-\frac{\kappa}{2}{p}+i\left(\Delta+g_{om}x-g\left|p\right|^2\right)+\sqrt{\kappa_{\text{ex}}}a_{\text{in}}  \\
m(\ddot{x}+\Gamma_m\dot{x}+\Omega_m^2x)=F_{\text{pol}} \\
\end{gathered} 
\end{equation}

with $\kappa$ the polariton mode loss rate, $\Delta=\omega_{\text{L}} - \omega_{\text{l}}$ the laser-polariton detuning, $g_\text{om}=g_{0}^{lp}/x_{\text{zpf}}$ with $x$ the mechanical displacement and $x_{\text{zpf}}$ the associated zero-point-fluctuation. $|p|^2$ is the intracavity polariton number, $a_\text{in}$ the optical field in the bus input waveguide, coupled at a rate $\kappa_{ex}$ to the polaritonic mode. The mechanical resonator dissipates energy at a rate $\Gamma_m$, and is subject to a polaritonic force $F_{\text{pol}}= \hbar g_{om} \times |p|^2$. 

The two coupled equations of \eqref{omPol} are however insufficient for our purposes, for two reasons. First, in semiconductor platforms, several works did report on the existence of an exciton reservoir coupling to the polariton fluid via dispersive Coulomb interactions at a rate $g_{R}$. This reservoir, at an energy close to that of polaritons, is crucial to understand the full dynamics of a resonantly driven polariton fluid ~\cite{Stepanov2019}. It can be included in the model via a simple rate equation on the reservoir population $n_{R}$, which relaxes at a rate $\kappa_{R}$ and is fed by a source term $\kappa_{I}\times |p|^2$, with $\kappa_{I}$ the capture rate into the reservoir. Second, in a real polaritonic device, thermal effects are often sizable and need to be included. Polaritons absorbed at a rate $\kappa_{\text{abs}}$ heat up the resonator by an amount $\Delta T$, attained after a relaxation time $\tau_{th}$, redshifting the polaritonic  frequency in proportion of a thermo-polaritonic coefficient $d\omega_{\text{l}}/dT$ and generating a photothermal force proportional to the temperature increase $F_{pth}=\alpha\Delta T$ \cite{Guha2017}. Taking into account all these effects, a minimal set of four coupled equations is necessary to depict the dynamics of a resonantly-driven optomechanical resonator filled by a polariton fluid:

\begin{equation}\label{FullSet}
\begin{gathered} 
\dot{p}=-\frac{\kappa}{2}{p}+i\left(\Delta+g_{om}x-g|p|^2- g_{R}n_{R} - \frac  {d\omega_{\mathrm{l}}}{dT}\Delta T\right)+\sqrt{\kappa_{\text{ex}}}a_{\mathrm{in}}  \\
\dot{n_{R}}=-\kappa_{R}n_{R}+\kappa_{I}\left|p\right|^2  \\
m(\ddot{x}+\Gamma_m\ \dot{x}+\Omega_m^2x)=F_{\mathrm{opt}}+F_{\mathrm{pth}} \\
\dot{\Delta}T=-\frac{1}{\tau_{\mathrm{th}}}\left(\Delta T-R_{\mathrm{th}}\kappa_{\mathrm{abs}}\hbar\omega_{\mathrm{l}}\left|p\right|^2\right) \\
\end{gathered} 
\end{equation}

with $R_{\text{th}}$ the resonator's thermal resistance. Below we will linearize these four coupled equations in order to obtain the linear optical response of the polariton optomechanical resonator under modulation of the input field $a_{\text{in}}$. Comparison with experiments will enable us to measure the polaritonic optomechanical coupling $g_0$, as well as all polaritonic nonlinearities.

\section{Experimental methods and analysis}

The employed disk resonators are fabricated in our clean-room \cite{De-Oliveira2022}. Their thickness is 200 nm and their diameter varies between 2 and 11 $\mu$m, see Fig. \ref{fig:presentation} (b). The starting epitaxial wafer is mainly composed of GaAs, within which five $12$ nm thick In$_{0.05}$Ga$_{0.95}$As/GaAs quantum wells were grown in stack, in order to support well-controlled excitons, see Fig. \ref{fig:presentation} (c) and \cite{De-Oliveira2022}. The disks are placed in a vacuum chamber with a pressure down to $10^{-7}$ mbar and cooled to 4K using a closed-cycle cryostat (beta-version of the Attocube Photonic Probe Station). For injecting and collecting light into/from the disk resonators, we use home-made conical optical fibers described in \cite{Pautrel:24}, with a total cone angle between 2 and 4 degrees. Cryogenic piezoelectric nanopositioners enable evanescent coupling between a junction of two such fibers and the disk (Fig. \ref{fig:presentation} (c)), with a low level of optical loss and low level of residual vibration \cite{Pautrel:24}. Two tunable continuous wave lasers are used in the polariton wavelength region: a mode-hop free laser diode (Velocity 930-945 nm) and a Ti:Sa laser (Solstis 700-1000 nm). With these laser spectroscopy tools, we measure WGM optical and polaritonic Qs evolving between $4\times10^{3}$ and $10^{4}$ at 4K. The mechanical Q of the fundamental Radial Breathing Mode (RBM) ranges from $500$ to $10,000$ at 4K, for the pedestal dimensions explored in this work \cite{APL2013}. In order to measure the linear coherent response of the polaritonic optomechanical resonator, the input laser power is modulated at frequency $\Omega$ using an electro-optic modulator. The modulator is fed by a vectorial network analyzer, which analyses the amplitude and phase response by demodulating the photocurrent proportional to the square of the optical output $a_{\text{out}}=a_{\text{in}}-\sqrt{\kappa_{\text{ex}}} p$.

\begin{figure}[ht]
\centering
\includegraphics[width=250pt]{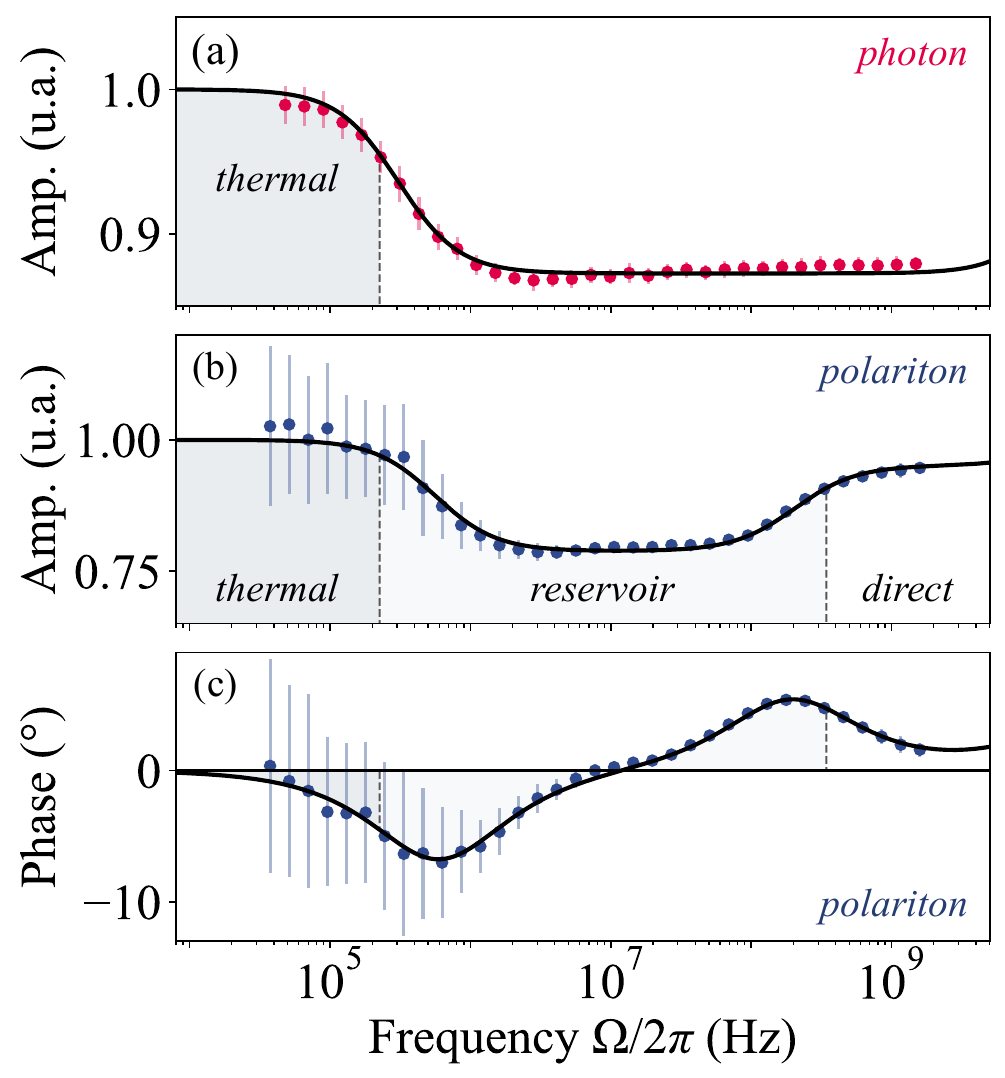}
\caption{\textbf{Modulation/demodulation experiments on a polaritonic disk resonator. (a)} Amplitude response on a photonic WGM. Experimental data (red dots) and mode (black lines).
\textbf{(b)} Amplitude response on a polaritonic WGM. Experimental data (blue dots) and mode (black lines).
\textbf{(c)} Phase response on the same polaritonic WGM. Experimental data (blue dots) and mode (black lines). All experiments are carried at 4 K, on a disk of radius 2 $\mu$m.
}
\label{fig:raw_data}
\end{figure}

Fig. \ref{fig:raw_data} presents the results of such modulation/demodulation experiments, carried by setting the laser resonant with either a photonic WGM (negligible excitonic weight)  or with a polaritonic WGM (sizable excitonic weight). To model these results, we linearize the four coupled equations of \eqref{FullSet} around a mean polariton field $\bar{p}$ to derive the linear response. We express the modulated input laser field as $\bar{a}_{\text{in}}+\delta a_{\text{in}}(t)=\bar{a}_{\text{in}}(1+\beta \cos{\Omega t})$ with $\beta \ll 1$, the polariton field as $\bar{p}+\delta p (t)$, and obtain the Fourier transform $\delta p(\Omega)=\int_{-\infty}^{\infty} \delta p(t)e^{-i\Omega t}dt$

\begin{equation}\hspace{-0.2cm}\label{pomega}
\delta p(\Omega)=i\sqrt{\kappa_{\text{ex}}}\frac{\beta}{2}\frac{(\bar{\Delta}+\Omega-i\frac{\kappa}{2})\bar{a}_{\text{in}}-(\tilde{g}_{LP}(\Omega)+\tilde{g}_{OM}(\Omega))\lvert\bar{p}\rvert^{2}(\bar{a}_{\text{in}}+\bar{a}_{\text{in}}^{*})} {\bar{\Delta}^{2}-2(\tilde{g}_{LP}(\Omega)+\tilde{g}_{OM}(\Omega))\bar{\Delta}\lvert\bar{p}\rvert^{2}-(\Omega-i\frac{\kappa}{2})^{2}},
\end{equation}
with a (frequency-dependent) effective polariton nonlinearity $\tilde{g}_{LP}$ that is the sum of the direct Coulomb nonlinearity $g$, of the exciton reservoir-mediated Coulomb nonlinearity, and of the thermo-polaritonic nonlinearity:
\begin{equation}\label{GLPomega}
\tilde{g}_{LP}(\Omega)=g+g_{R}\kappa_{I}\tau_{R}\frac{1}{1+i\Omega\tau_{R}}+g_{th}\frac{1}{1+i\Omega\tau_{th}},
\end{equation}
where $g_{th}=R_{\text{th}}\kappa_{\text{abs}}\hbar\omega_{\text{L}}\times d\omega_{\text{l}}/dT$. A (frequency-dependent) optomechanical polariton nonlinearity $\tilde{g}_{OM}$ appears as well in the response \eqref{pomega} : 
\begin{equation}\label{GOMomega}
\tilde{g}_{OM}(\Omega)=-2\Omega_m \frac{ g_{0}^{2} + g_{0} \alpha /(1+i\Omega\tau_{th}) }  { \Omega_m^{2}-\Omega^{2}+i\Gamma_m\Omega },
\end{equation}
which embeds the resonant mechanical response to the polaritonic and photothermal forces. The effective detuning at mean-field is $\bar{\Delta}=\Delta-\tilde{g}_{LP}(0)\lvert\bar{p}\rvert^{2}$, enabling to finally express the demodulated signal $\lvert a_{\text{out}}\rvert^{2} (\Omega)$. In Fig. \ref{fig:raw_data}, where the data are displayed over a frequency range ($10^{4}-10^{9}$ Hz) that is orders of magnitude larger than the mechanical line width $\Gamma_m \leq 1$ MHz, the mechanical resonance is invisible. We can hence set $\tilde{g}_{OM}=0$ in our model to obtain the fits displayed as solid lines. The fits agree remarkably well with experimental data, all over the explored frequency range, be it for the amplitude or the phase. This all demonstrates that we have an experimental method and a model to investigate the frequency response of polaritonic optomechanical resonators. We will use these tools below to measure and quantify both polaritonic nonlinearities and the resonant optomechanical response involving polaritonic effects.

\section{Polariton nonlinearities}

% As illustrated in Figure 2, where the data are displayed over a frequency range (104 −109 Hz), the data display facilitates the distinct observation of the three polariton nonlinearities, a capability that was not previously available in first experiments measuring polariton nonlinearities. To characterize these nonlinearities, we can make the assumption of g_(OM) = 0 in our model to obtain the fits displayed as solid lines. This is because the mechanical linewidth is very small compared to the range (< 1 MHz). 

The first subject we want to address with our method is that of polaritonic nonlinearities. This is indeed a central concern in the polaritonics community, where reported values for the nonlinearity did vary by orders of magnitude since early reports \cite{Amo2009}. This state of fact has challenged the simple picture of polariton-polariton interactions being solely governed by Coulomb interactions between their excitonic parts. Our method opens a new window on this topic. For the photonic WGM, Fig. \ref{fig:raw_data} (a) shows an amplitude response with a single-frequency cut-off around $10^{5}$ Hz, which reaches a flat response at higher frequency. In that case where the excitonic fraction is negligible, we can set $g=g_{R}=0$. Since $\kappa$, $\kappa_{\text{ex}}$, $\bar{\Delta}$ and $\lvert\bar{p}\rvert^{2}$ are measured independently, the fit by our model enables determining both $g_{th}$ and $\tau_{th}$, retrieving the expressions and conclusions found for a photonic WGM in \cite{Sbarra2021}. While the determined $g_{th}$ depends on the specific considered photonic WGM, $\tau_{th}$ only depends on the disk geometry and holds the same value for photonic and polaritonic WGMs. For the polaritonic WGM, Fig. \ref{fig:raw_data} (b) shows a richer amplitude response with a supplementary turning frequency at a few $10^{8}$ Hz, above which the amplitude re-increases. The phase response of Fig. \ref{fig:raw_data} (c) shows the changes of sign associated to the two turning frequencies. Our model correctly predicts this frequency evolution, and the fit to data enables obtaining a unique value for the 3 supplementary model parameters $g$, $g_{R}\kappa_{I}\tau_{R}$ and $\tau_{R}$. Ultimately, since the DC laser spectroscopy of the polariton resonance fixes the quantity $g+g_{R}\kappa_{I}\tau_{R}+g_{th}$, we can determine the amplitude of all nonlinearities and timescales. The second turning frequency appears to be determined by the value of $\tau_{R}$, leading to the following interpretation. At the highest frequencies (above $1$ GHz) the sole direct Coulomb nonlinearity between polaritons contributes to the frequency response. At intermediate frequencies (between $1$ MHz and $1$ GHz), the exciton reservoir-mediated nonlinearity adds its contribution, with the same sign as the direct nonlinearity ($g, g_{R}\kappa_{I}\tau_{R}>0$). At the smallest frequencies (below $1$ MHz), the thermal polaritonic nonlinearity adds its contribution, with a sign that is however opposite to the two others ($g_{th}<0$). This interpretation is illustrated in Fig. \ref{fig:raw_data} (b).

\begin{figure}[]
\centering
\includegraphics[width=230pt]{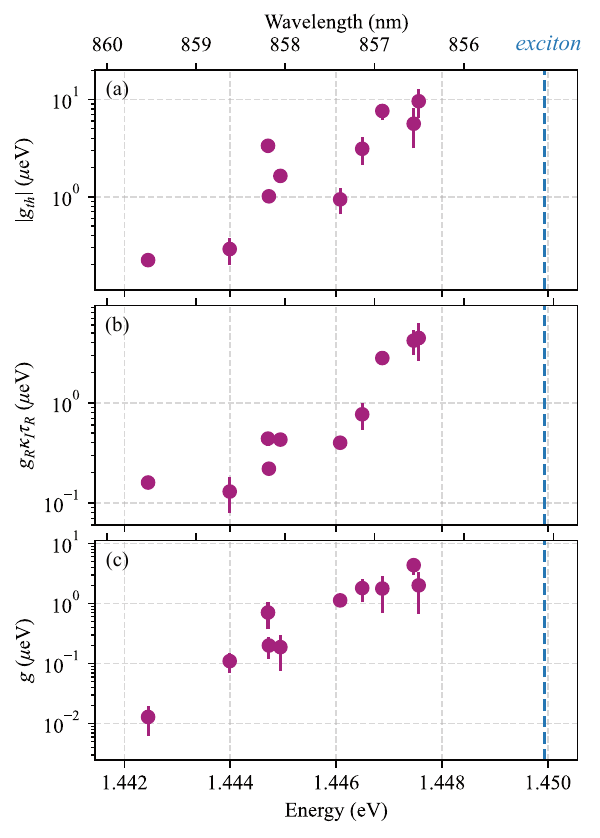}
\caption{\textbf{Polaritonic nonlinearities as function of polariton energy}. Norm of the thermal ($g_{th}$, \textbf{(a)}), exciton reservoir-mediated ($g_{R}\kappa_{I}\tau_{R}$, \textbf{(b)}) and direct ($g$, \textbf{(c)}) polaritonic nonlinearity, as function of the energy of the investigated WGM polaritonic mode. The disk radius is 2 $\mu$m. The temperature is 4 K.}
\label{fig:nonlinear}
\end{figure}

This overall frequency response was measured consistently on all polaritonic modes investigated in this work, enabling a systematic extraction of the value of the three involved nonlinearities and timescales. Figure \ref{fig:nonlinear} shows the amplitude of the three nonlinearities as function of the energy of the polariton mode. The three nonlinearities show comparable amplitude, however with a clear hierarchy ($\lvert g_{th}\rvert \ge g_{R}\kappa_{I}\tau_{R} \ge g$) that seems to prevail for all polaritonic modes. The nonlinearity amplitudes increase as the polariton energy approaches the exciton energy (hence as the excitonic fraction increases), evolving from $10^{-2}$ to $10^{1}$ \micro eV. This sharp increase is the strong indication of the role played by the exciton in all these nonlinear processes. This is expected for the direct polariton nonlinearity, which relies on Coulomb interactions between the excitonic fractions, but less documented for the two other types of nonlinearity, which have rarely been measured. The exciton reservoir mediated nonlinearity being also Coulombian in nature, an increase with the exciton fraction is there as well consistent. For the thermal nonlinearity, we interpret that the increase is mainly governed by an increase of $d\omega_{\text{l}}/dT$ with the exciton fraction, since we observed a moderate increase (less than a factor 3) of the absorption rate $\kappa_{abs}$ over the same range (not shown). Our approach for measuring nonlinearities being modal and calibrated, it is instructive to compare it to recent works that tried to individualize the direct Coulomb nonlinearity by other means ~\cite{Estrecho2019, Schnuriger2026}. The values we find for the direct nonlinearity, once multiplied by an effective WGM area and normalized by the number of quantum wells, come close to those reported in \cite{Schnuriger}. For example, we estimate a value between 2 and 8 \micro eV.\micro m $^2$ for an excitonic fraction $50\%<X^{2}<75\%$ and between $0.1$ and $1$ $\micro eV. \micro m^2$ for an excitonic fraction $20\%<X^{2}<40\%$. Regarding timescales, we systematically measured $\tau_{R}$ and $\tau_{th}$ for all investigated polariton modes, and observed that they barely vary with the excitonic fraction (not shown). $\tau_{R}$ is comprised between $0.5$ and $1.5$ ns, a typical range for exciton relaxation ~\cite{Stepanov2019}, while $\tau_{th}$ evolves between $0.5$ and $1.5$ $\micro$s, consistent with numerical thermal simulations of the disk resonator ~\cite{Guha2017}. Overall, our method provides us with a complete and quantitative picture of polariton nonlinearities in a semiconductor resonator. This picture, with three distinct nonlinearities competing on distinct timescales, sheds new light on the large data spread found in the literature, with reported nonlinearities that vary by almost three decades (see discussion in \cite{Schnuriger}). Given the ongoing debates in the community regarding the various mechanisms underlying polariton nonlinearities~\cite{Richard2026, Christensen2024} our methods opens exciting perspectives for further investigations of microscopic models of polariton-polariton interactions.

\section{Polaritonic force and $g_{0}$ boost}

\begin{figure}[ht]
\centering
\includegraphics[width=240pt]{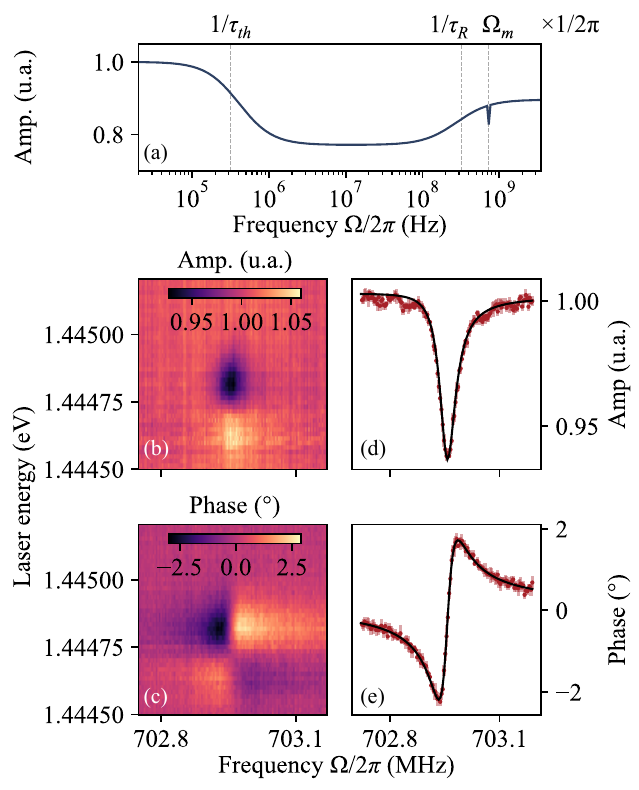}
\caption{\textbf{Resonant optomechanical response of the polaritonic resonator}. \textbf{(a)} Illustration of the mechanical resonance superposing over the broad frequency response of the polaritonic resonator. \textbf{(b)} Amplitude of the frequency response around the RBM1 mechanical resonance, for varying input laser energy. \textbf{(c)} Phase response, for varying input laser energy. \textbf{(d)} Amplitude of the frequency response, for a fixed input laser energy of $1.444869$ eV, as function of the modulation frequency. \textbf{(e)} Phase of the frequency response, for the same fixed input laser energy. Disk radius is 2 $\mu$m. Temperature is 4 K.}
\label{fig:mechanical}
\end{figure}

The second domain where we are eager to apply our methods is that of polariton-induced forces and optomechanical coupling. Measuring the resonant polaritonic forces in a concrete device has so far remained a challenge, and polariton-assisted $g_{0}$ were not directly measured either. Our method breaks open these two locking points. To that purpose, after having analyzed the response of the polaritonic resonator with a low frequency resolution and over a broad frequency range ($10^{4}-10^{9}$ Hz), we now turn our attention to the response close to the disk mechanical resonance, which sits around $\Omega_{m}/2\pi\simeq700$ MHz (see Fig. \ref{fig:mechanical}.(a)). To that purpose, we increase the frequency resolution and reduce the frequency span in order to reveal the details of the amplitude and phase response within the mechanical line width smaller than 1 MHz). Fig. \ref{fig:mechanical}. (b) and (c) show the detailed amplitude and phase response close to $\Omega_{m}/2\pi$, for a laser energy varying in the vicinity of the resonator polaritonic resonance, hence for a varying detuning $\Delta$. As apparent in these plots, the sign of the resonant response depends on the chosen detuning. Fig. \ref{fig:mechanical}. (d) and (e) show the response for a specific detuning, corresponding to a laser energy of $1.444869$ eV. While the resonant response adopts in general a Fano-like shape, for this specific detuning it shows a quasi-Lorentzian response, reminiscent of the underlying harmonic mechanical resonator. It can be interpreted as the resonant mechanical response of the disk to polaritonic and photothermal forces, which through optomechanical coupling impacts in return the response of the nonlinear polaritonic resonator. If properly modeled, this spectral feature provides access to a calibrated measurement of polaritonic and photothermal forces. In order to model this resonant response, we need now to consider $\tilde{g}_{OM} \neq 0$. With respect to the fitting procedure of the prior section, Eq. \eqref{GOMomega} now introduces two supplementary parameters: $\alpha$ and $g_{0}$. Fitting the resonant response for varying detunings $\Delta$ leads to a unique and unambiguous determination of $\alpha$ and $g_{0}$, for each polaritonic mode. Fig. \ref{fig:mechanical}. (d) and (e) show the result of this fitting procedure: the model (black lines) perfectly reproduces the measurement.

\begin{figure*}
    \centering
    \includegraphics[width=340pt]{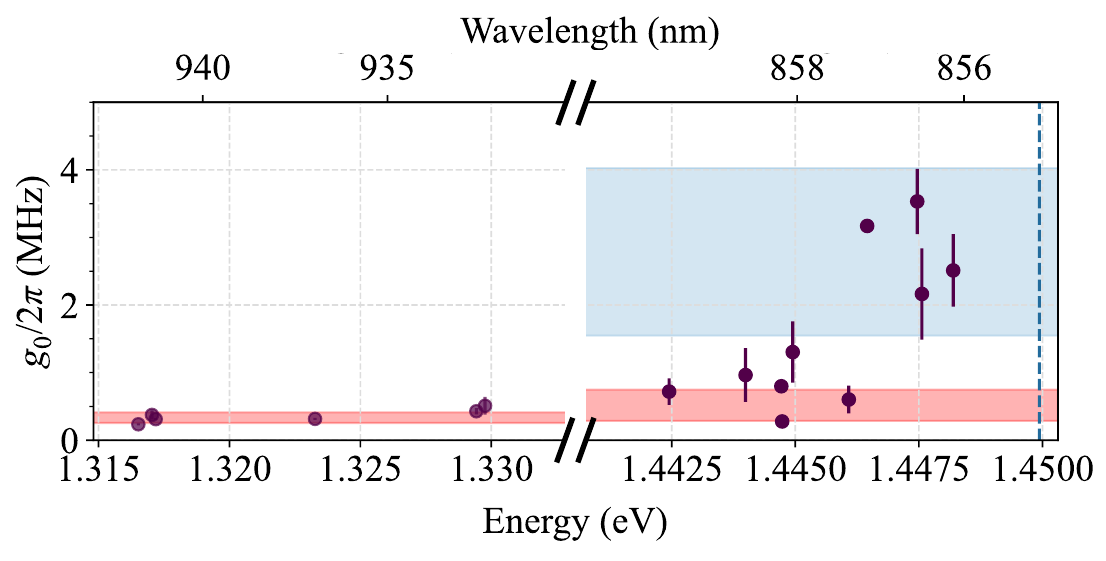}
    \caption{\textbf{Polaritonic boost of $g_0$}. $g_0/2\pi$ as a function of the WGM energy. The purple disks represent values extracted from the measurement, with their error bar. The red strips correspond to simulated values of the photonic $g_{0}^{opt}$, while the blue strip corresponds to the excitonic $g_{0}^{exc}$. Measurements and simulations are for a 2 $\mu$m radius disk at 4 K.}
    \label{fig5}
\end{figure*}

Being now equipped with an experimental method to measure $g_{0}$ that is valid for both photonic and polaritonic modes, we can turn our attention to a comparison between both classes of modes. To that purpose, in Fig. \ref{fig5} we plot the values of $g_0$ for all WGM investigated in this work, as a function of the WGM energy. The investigation starts with WGM in the wavelength region 930-945 nm ($1.315$$<$$E$$<$$1.33$ eV), spectrally far enough from the exciton to be considered as mere photonic WGMs. It then proceeds with a monotonous increase of the WGM energy, entering the polaritonic WGM range ($1.44$$<$$E$$<$$1.45$ eV) as the energy progressively approaches that of the exciton ($1.45$ eV). We observe a clear monotonous increase of $g_{0}$ as we move from the bare photonic to the polaritonic regime: $g_{0}/2\pi$ starts at $0.2$ MHz in photonic-like WGMs, and culminates close to $4$ MHz for the most excitonic polaritons. This represents more than a decade of improvement in $g_{0}$ with respect to the state of the art in WGM optomechanical structures \cite{Ding2011apl,Favero2014}, and constitutes a direct experimental proof of the polaritonic boost of the optomechanical coupling.

Let us try now to analyze and model quantitatively these results. A first remark is that the photoelastic couplings increase as we come closer to the host material bandgap ($E=1.515$ eV for GaAs at $4$ K) \cite{Feldman1968Piezobirefringence, Renosi1993}. Since $g_{0}$ strongly depends on photoelasticity, we must first quantitatively assess this contribution to the $g_{0}$ increase. To that purpose, we use FEM simulations of $g_{0}^{opt}$ that embed both photoelastic and moving boundary effects \cite{Favero2014}, considering the first Radial Breathing Mode (RBM1) as mechanical mode. As an exact identification of the radial and azimuthal numbers $p$ and $m$ for each measured WGM polaritonic resonance is challenging, we run these simulations for an ensemble of $p$ and $m$ corresponding to WGMs in the same spectral region as the resonance. The photoelastic coefficients are extracted from the literature at room temperature \cite{Renosi1993} and shifted to 4 K following the energy gap evolution. This approach enables a numerical computation of the photonic coupling $g_{0}^{opt}$ for each spectral region (photonic region: $1.315-1.330$ eV and polaritonic region: $1.44-1.45$ $eV$). The uncertainty in the simulation is the convolution of uncertainties on $p$ and $m$, and on photoelastic coefficients. The simulated $g_{0}^{opt}$ and its uncertainty are represented by an horizontal light red strip in Fig. \ref{fig5}, for each spectral region. For the photonic WGM spectral region, we observe that the numerically simulated photonic coupling $g_{0}^{opt}$ agrees very well with experimental measurements. In the polaritonic WGM spectral region in contrast, the measured $g_{0}$ values are close to a decade larger than the numerically simulated value for $g_{0}^{opt}$. This demonstrates that the increase of $g_{0}$ in this wavelength region is mostly due to polaritonic effects, and not to the increase of photoelasticity of the host material.

To go further, we turn back our attention to Eq. \eqref{eq:g_0pol}, which involves the photonic optomechanical coupling $g_{0}^{opt}$ and the vacuum exciton-phonon coupling $g_{0}^{exc}$. In \cite{De-Oliveira2022,CarlonZambon2022}, we established methods to compute $g_{0}^{exc}$. In a perfect disk, excitonic modes wavefunctions adopt a WGM structure parametrized by radial and azimuthal numbers $p$ and $m$. $g_{0}^{exc}$ is expressed for a single quantum well inserted at the vertical coordinate $z_{QW}$:

\begin{equation}\label{excitonic}
\hbar g_0^{exc} =-\left( a_e - a_h \right)\frac{x_{\mathrm{zpf}}}{|\mathbf{u}_{max}|} \int_{S} |F^{m,p}_\parallel (\mathbf{R}_\parallel)|^2 \mathbf{\nabla}.\mathbf{u}(\mathbf{R}_\parallel,z_{QW}) d\mathbf{R}_\parallel
\end{equation}

where $a_e$ and $a_h$ are the electron and hole deformation potentials in InGaAs \cite{Vurgaftman2001AlloysProperties, De-Oliveira2022}, $\mathbf{u}_{max}$ is the displacement at the reduction point chosen for the mechanical mode, which by convention is here a point of maximal displacement. $x_{\mathrm{ zpf}}=\sqrt{\frac{\hbar}{2m_{\mathrm{eff}}\Omega_{m}}}$ with $m_{\mathrm{eff}}$ the effective mass calculated with the reduction point. $F_\parallel$ is the normalized in-plane envelope wave function of the quantum well exciton \cite{De-Oliveira2022}, $\mathbf{R_\parallel}$ the in-plane radial vector, $\mathbf{u}$ the displacement vector associated to the considered mechanical mode (here the first RBM), which can be computed by FEM simulations. As our resonators embody five quantum wells and we are using the bright (symmetric) exciton mode distributed over the five wells, $g_{0}^{exc}$ is increased by a factor $\sqrt{5}$ with respect to Eq. \eqref{excitonic} \cite{De-Oliveira2022}. We employed this calculation method to compute $g_{0}^{exc}$ for all $(p,m)$ numbers considered in the polaritonic spectral region, obtaining the interval of values represented by the light blue strip in Fig. \ref{fig5}. We observe that our largest measured $g_{0}$ values precisely fall within this interval, which is consistent with Eq. \eqref{eq:g_0pol} that shows that for the most excitonic polaritons $g_{0}$ approaches $g_{0}^{exc}$. Our measurements also point to a category of modes where $g_{0}$ sits between $g_{0}^{opt}$ (red strip) and $g_{0}^{exc}$ (blue strip), again consistent with Eq. \eqref{eq:g_0pol} for polaritons having more balanced photonic and excitonic weights. Overall, the measurements and model results reported in Fig. \ref{fig5} constitute an unambiguous proof of the polaritonic boost of $g_{0}$ and of the validity ot Eq. \eqref{eq:g_0pol}.

\section{Conclusion}

In this work, we experimentally demonstrated that the insertion of polaritons in an optomechanical resonator leads to a boosted optomechanics with nonlinear photons. While experimentally grounding prior theoretical works, our investigation also points towards the importance of including an exciton reservoir and thermal effects to correctly depict the dynamics of such a system. This led us to propose a minimal set of four constitutive equations, which we did linearize in order to express the linear frequency response of a polaritonic optomechanical resonator. Our resonant measurement of such frequency response, carried in a systematic manner on various polaritonic modes, allowed us extracting the optomechanical coupling $g_{0}$ and all polaritonic nonlinearities. For the most excitonic polaritons $g_{0}$ shows a boost of more than a decade, a result perfectly in line with our theoretical expectations. While already leading to record $g_{0}$ values for whispering gallery structures in this report, further optimization is in principle possible. Eq. \eqref{excitonic} is nothing but the expression of a modal overlap, and better spatial matching of the exciton and mechanical wave functions will enable other decade-scale gains in $g_{0}$, getting closer to the regime of single photon optomechanical effects \cite{Nunnenkamp11,Rabl2010}. At the same time, three distinct and commensurate nonlinearities are revealed by our optomechanically inspired measurements, an aspect that will be of utmost importance for further investigations of quantum fluids of polaritons. 

The present paper establishes a proper analysis frame for semiconductor polaritonic optomechanics. At the same time, it validates a reference experimental platform for optomechanics with nonlinear photons, of direct interest to other communities facing the same Hamiltonian, like researchers parametrically coupling mechanical and magnetic degrees of freedom \cite{Kirchmair23}. Amongst interesting effects at the intersection of nonlinear photons and optomechanics, one finds the possibility of enhanced optical cooling or amplification of mechanical motion \cite{CarlonZambon2022,Metelmann25}, which might enable obtaining low-power optomechanical oscillators, or motional quantum state preparation with alleviated optical power requirements. But beyond this, the present work established a very natural path between the mechanical motion of solid bodies and quantum fluids of light \cite{Carusotto15}. Polariton fluids notably experience transition to superfluidity \cite{Amo2009}, an aspect that could directly translate into the optomechanical response of the resonator. With a complete model now established, a tangible next step would be to optomechanically witness this transition to light superfluidity.

\section{Acknowledgements}
We acknowledge M. Nicolas and B. Janvier for the conception and fabrication of fiber holders, P. Filloux for assistance with the fabrication of samples, R. de Oliveira and Z. Denis for help in simulations of exciton-phonon couplings. The research was supported by the European Research Council via the NOMLI project (770933).

\medskip

\bibliography{sample}

\end{document}